\documentclass[12pt]{article}
\usepackage{graphics, setspace,cite}
\usepackage{mathrsfs}
\usepackage{amsfonts}
\usepackage{textcomp}
\usepackage{color}
\usepackage{amsmath}
\usepackage{graphicx}
\usepackage{amssymb}
\usepackage{authblk}

\usepackage{epstopdf}
\DeclareGraphicsExtensions{.eps}

\begin{document}
\thispagestyle{empty}

\def\theequation{\arabic{section}.\arabic{equation}}
\def\a{\alpha}
\def\b{\beta}
\def\g{\gamma}
\def\d{\delta}
\def\dd{\rm d}
\def\e{\epsilon}
\def\ve{\varepsilon}
\def\z{\zeta}
\def\B{\mbox{\bf B}}\def\cp{\mathbb {CP}^3}

\newcommand{\h}{\hspace{0.5cm}}

\begin{titlepage}

\renewcommand{\thefootnote}{\fnsymbol{footnote}}
\begin{center}
{\Large \bf The exact solution of a  generalized two-spins model}
\end{center}
\vskip 1.2cm \centerline{\bf G. Santos}

\vskip 10mm
\centerline{\sl Universidade Federal de Sergipe - UFS} 
\centerline{Centro de Ci\^encias Exatas e Tecnologia - CCET} 
\centerline{Departamento de F{\'i}sica}
\centerline{\sl Cidade Univ. Prof. Jos\'e Alo{\'i}sio de Campos} 
\centerline{\sl Av. Marechal Rondon, s/n, Jd. Rosa Elze}
\centerline{\sl S\~ao Crist\'ov\~ao/SE - Brazil.}
\vskip .5cm

\vspace*{0.6cm} \centerline{\tt gnsfilho@gmail.com}

\vskip 20mm

\baselineskip 18pt

\begin{center}
{\bf Abstract}
\end{center}

We present the exact solution of a family of two-spins models. The models are solved by the algebraic Bethe ansatz method using the $gl(2)$-invariant $R$-matrix and a multi-spins Lax operator. The interactions are by the Heisenberg spins exchange. We are also considering  magnetic $B$-fields and a term for Haldane single spin anisotropy.

\end{titlepage}
\newpage
\baselineskip 18pt

\def\nn{\nonumber}
\def\tr{{\rm tr}\,}
\def\p{\partial}
\newcommand{\non}{\nonumber}
\newcommand{\bea}{\begin{eqnarray}}
\newcommand{\eea}{\end{eqnarray}}
\newcommand{\bde}{{\bf e}}
\renewcommand{\thefootnote}{\fnsymbol{footnote}}
\newcommand{\be}{\begin{eqnarray}}
\newcommand{\ee}{\end{eqnarray}}

\vskip 0cm

\renewcommand{\thefootnote}{\arabic{footnote}}
\setcounter{footnote}{0}

\setcounter{equation}{0}
\section{Introduction}

   The algebraic formulation of the Bethe ansatz method, and the associated quantum inverse scattering method (QISM), was primarily  developed in \cite{fst,ks,takhtajan,korepin,faddeev} and it has been used to study a considerable number of exactly solvable systems, such as, one-dimensional spin chains, quantum field theory of one-dimensional 
interacting bosons \cite{korepin1} and fermions \cite{yang}, two-dimensional lattice models \cite{korepin2}, systems of strongly correlated electrons \cite{ek,ek2}, conformal field theory \cite{blz}, integrable systems in high energy physics \cite{lipatov, korch, belitsky,Gromov1,Gromov2,CAhn}  and quantum algebras  \cite{jimbo85,jimbo86,drinfeld,frt}.  More recently solvable models have also shown up in relation to string theories \cite{Dorey,Zarembo,Plefka}.  Remarkably it is important to mention that exactly solvable models are recently finding their way into the lab, mainly in the context of ultracold atoms \cite{batchelor2,kino1,kino2,kitagawa,haller,liao,coldea} but also in nuclear magnetic resonance (NMR) experiments\cite{screp,nmr1} 
becoming its study as well as the derivation of 
new models an even more fascinating field. We are considering in this work a multi-spins Lax operator  that like the multi-bosons Lax operator in  \cite{multistates} permits to solve a family of models, helping to increase the number of integrable models solved by the algebraic Bethe ansatz method. These spins cluster integrable models are  important as for example in quantum computation and quantum information,  quantum simulation using nuclear magnetic resonance (NMR) \cite{Roditi1,Roditi2,Ivan} or trapped ions \cite{Lanyon} as well as in molecular magnetism \cite{LNP645}.  Recently, spins configurations like these were proposed in    \cite{Lanyon} to a universal digital quantum simulation with trapped ions and an experimental realization of the Yang-Baxter equation was realized via NMR interferometry \cite{screp} using   Iodotrifluoroethylene $(C_2F_3I)$ molecules. Another important application of this spins configurations is as a spin network that appears in quantum gravity \cite{QG1,QG2}.

The paper is organized as follows. In section 2, I will describe briefly the algebraic
Bethe ansatz method and present the multi-spins Lax operators. In section 3, I present a generalized two-spins model. In section 4, I present the exact solution. In section 5, I summarize the results.

\section{Algebraic Bethe ansatz method}

In this section we will shortly describe the algebraic Bethe ansatz method and present the Lax operators used to get the solution of the 
models \cite{jonjpa,Roditi}. We begin with the $gl(2)$-invariant $R$-matrix, depending on the spectral parameter $u$,

\begin{equation}
R(u)= \left( \begin{array}{cccc}
1 & 0 & 0 & 0\\
0 & b(u) & c(u) & 0\\
0 & c(u) & b(u) & 0\\
0 & 0 & 0 & 1\end{array}\right),\end{equation}

\noindent with $b(u)=u/(u+\eta)$, $c(u)=\eta/(u+\eta)$ and $b(u) + c(u) = 1$. Above,
$\eta$ is an arbitrary parameter, to be chosen later.

It is easy to check that $R(u)$ satisfies the Yang-Baxter equation

\begin{equation}
R_{12}(u-v)R_{13}(u)R_{23}(v)=R_{23}(v)R_{13}(u)R_{12}(u-v),
\end{equation}

\noindent where $R_{jk}(u)$ denotes the matrix acting non-trivially
on the $j$-th and the $k$-th spaces and as the identity on the remaining
space.

Next we define the monodromy matrix  $\hat{T}(u)$,

\begin{equation}
\hat{T}(u)= \left( \begin{array}{cc}
 \hat{A}(u) & \hat{B}(u)\\
 \hat{C}(u) & \hat{D}(u)\end{array}\right),\label{monod}
\end{equation}

\noindent such that the Yang-Baxter algebra is satisfied

\begin{equation}
R_{12}(u-v)\hat{T}_{1}(u)\hat{T}_{2}(v) = \hat{T}_{2}(v)\hat{T}_{1}(u)R_{12}(u-v).\label{RTT}
\end{equation}

\noindent In what follows we will choose a realization for the monodromy matrix $\pi(\hat{T}(u))=\hat{L}(u)$  
to obtain the solutions of a family of two-spins models. In this construction, the Lax operators $\hat{L}(u)$  have to satisfy the relation

\begin{equation}
R_{12}(u-v)\hat{L}_{1}(u)\hat{L}_{2}(v)=\hat{L}_{2}(v)\hat{L}_{1}(u)R_{12}(u-v).
\label{RLL}
\end{equation}

Then, defining the transfer matrix, as usual, through

\begin{equation}
\hat{t}(u)= tr \;\pi(\hat{T}(u)) = \pi(\hat{A}(u) + \hat{D}(u)),
\label{trTu}
\end{equation}
\noindent it follows from (\ref{RTT}) that the transfer matrix commutes for
different values of the spectral parameter; i. e.,

\begin{equation}
[\hat{t}(u),\hat{t}(v)]=0, \;\;\;\;\;\;\; \forall \;u,\;v.
\end{equation}
\noindent Consequently, the models derived from this transfer matrix will be integrable. Another consequence is that the 
coefficients $\hat{\mathcal{C}}_k$ in the transfer matrix $\hat{t}(u)$,

\begin{equation}
\hat{t}(u) = \sum_{k} \hat{\mathcal{C}}_k u^k,
\end{equation}
\noindent are conserved quantities or simply $c$-numbers, with

\begin{equation}
[\hat{\mathcal{C}}_j,\hat{\mathcal{C}}_k] = 0, \;\;\;\;\;\;\; \forall \;j,\;k.
\label{invol}
\end{equation}

If the transfer matrix $\hat{t}(u)$ is a polynomial function in $u$, with $k \geq 0$, it is easy to see that,

\begin{equation}
\hat{\mathcal{C}}_0 = \hat{t}(0) \;\;\; \mbox{and} \;\;\; \hat{\mathcal{C}}_k = \frac{1}{k!}\left.\frac{d^k\hat{t}(u)}{du^k}\right|_{u=0}. 
\label{C14b}
\end{equation}

 For the standard $SU(2)$ spin operators satisfying the commutation relations 

\begin{equation}
 [\hat{S}_{pj}^{+},\hat{S}_{qk}^{-}] = 2 \; \delta_{jk} \; \delta_{pq}\;\hat{S}_{kq}^{z}, \qquad [\hat{S}_{pj}^{z},\hat{S}_{qk}^{\pm}] = \pm  \; \delta_{jk} \; \delta_{pq}\;\hat{S}_{kq}^{\pm},
\label{su2a} 
\end{equation}

\begin{equation}
[\hat{\vec{S}}_{pj}^{2},\hat{S}_{qk}^{\pm}] = 0, 
\label{su2b} 
\end{equation}
\noindent with $p,q = a \;\mbox{or}\; b$, $j = 1,\ldots, n$ and $k = 1, \ldots, m$, we have the following Lax operators,

\begin{equation}
\hat{L}^{S_a^n}(u)=
\left(\begin{array}{cc}
\gamma_a u\hat{I} -  \eta \sum_{j=1}^n \gamma_a\hat{S}_{aj}^z & -\eta \sum_{j=1}^n \alpha_{aj}\hat{S}_{aj}^{+} \\
-\eta \sum_{j=1}^n \beta_{aj}\hat{S}_{aj}^{-} & \rho_a u\hat{I} +  \eta \sum_{j=1}^n \rho_a\hat{S}_{aj}^z
\end{array}\right),
\label{Laxm1}
\end{equation}

\begin{equation}
\hat{L}^{S_b^m}(u)=
\left(\begin{array}{cc}
\gamma_b u\hat{I} -  \eta \sum_{k=1}^m \gamma_b\hat{S}_{bk}^z & -\eta \sum_{k=1}^m \alpha_{bk}\hat{S}_{bk}^{+} \\
-\eta \sum_{k=1}^m \beta_{bk} \hat{S}_{bk}^{-} & \rho_b u\hat{I} +  \eta \sum_{k=1}^m \rho_b\hat{S}_{bk}^z
\end{array}\right),
\label{Laxm2}
\end{equation}
\noindent with the condition $\alpha_{aj}\beta_{aj}=\gamma_a\rho_a$ and $\alpha_{bk}\beta_{bk}=\gamma_b\rho_b, \;\; \forall \;j,\;k$. The above Lax operators satisfies the equation (\ref{RLL}).

\section{Generalized two-spins model}
In this section I present the generalized two-spins model with two different spins: $a$ and $b$. The  Hamiltonian is,

\begin{eqnarray}
\hat{H} & = & \sum_{j=1}^n  B^z_{aj}\hat{S}^z_{aj} + \sum_{k=1}^m  B^z_{bk}\hat{S}^z_{bk} +   \sum_{j=1}^n D_{aj} (\hat{S}_{aj}^z)^2 +  \sum_{k=1}^m  D_{bk} (\hat{S}^z_{bk})^2 \nonumber \\&+& \sum_{i=1}^n\sum_{j\neq i}^n  J^z_{aiaj} \hat{S}^z_{ai}\hat{S}^z_{aj} + \sum_{k=1}^m\sum_{l\neq k}^m  J^z_{bkbl} \hat{S}^z_{bk}\hat{S}^z_{bl} \nonumber \\ 
&+& \sum_{j=1}^n\sum_{k=1}^m  J^z_{ajbk} \hat{S}^z_{aj}\hat{S}^z_{bk} + \frac{1}{2}\sum_{j=1}^n\sum_{k=1}^m  J^{xy}_{ajbk} (\hat{S}^{+}_{aj}\hat{S}^{-}_{bk} + \hat{S}^{-}_{aj}\hat{S}^{+}_{bk}).
\label{H1spin}
\end{eqnarray}
\noindent  The parameters $B^z_{aj}$ and $B^z_{bk}$ are magnetic fields in the $z$-direction, $J^z_{aiaj}$ are  the exchange interaction parameters between the spins $a$ and $J^z_{bkbl}$ are  the exchange interaction parameters between the spins  $b$, $J^z_{ajbk}$ are the exchange interaction parameters between the spins $a$ and $b$ in the $z$-direction and $J^{xy}_{ajbk}\equiv J^{x}_{ajbk} = J^{y}_{ajbk}$ are the exchange interaction parameters between the spins $aj$ and  $bk$ in the $x$ and $y$-direction, and $D_{aj}$ and $D_{bk}$ are the Haldane single spin anisotropy  parameters for the spins $aj$ and for the spins $bk$, respectively \cite{LNP645}.

The spins operators satisfies the $O(3)$ algebra, with the following commutation relations

\begin{equation}
[\hat{S}^r_{pj},\hat{S}^s_{qk}] = i\hbar\delta_{pq}\delta_{jk}\varepsilon_{rst} \hat{S}^t_{qk},
\label{ccrs1}
\end{equation}
\noindent with $\varepsilon_{rst}$ the completely antisymmetric Levi-Civita tensor, $p,q = a,\; b$ the two spins labels, $r,s,t = x, y , z$ the spin components and $j=1\ldots n$, $k=1\ldots m$.  Using the rising and the lowing operators $\hat{S}^{\pm}$  we get the commutation relations in (\ref{su2a}) e (\ref{su2b})

\begin{equation}
\hat{S}^{\pm}_{pj} = \hat{S}^x_{pj} \pm i \hat{S}^y_{pj}.
\end{equation}

The total spin is 

\begin{equation}
\hat{\vec{S}}_T = \sum_{j=1}^n \hat{\vec{S}}_{aj} + \sum_{k=1}^m \hat{\vec{S}}_{bk},
\end{equation} 
\noindent with

\begin{equation}
\hat{\vec{S}}_T^2 = \sum_{i,j=1}^n \hat{\vec{S}}_{ai}\cdot \hat{\vec{S}}_{aj} +  \sum_{k,l=1}^m \hat{\vec{S}}_{bk}\cdot \hat{\vec{S}}_{bl} + 2 \sum_{i=x,y,z}\sum_{j=1}^n\sum_{k=1}^m \hat{S}^i_{aj} \hat{S}^i_{bk}.
\end{equation}

The $z$-component of the total spin operator, $\hat{S}_T^z$,

\begin{equation}
\hat{S}_T^z = \sum_{j=1}^n \hat{S}^z_{aj} + \sum_{k=1}^m \hat{S}^z_{bk},
\end{equation} 
\noindent with

\begin{equation}
(\hat{S}_T^z)^2 = \sum_{i,j=1}^n \hat{S}^z_{ai} \hat{S}^z_{aj} +  \sum_{k,l=1}^m \hat{S}^z_{bk} \hat{S}^z_{bl} + 2 \sum_{j=1}^n\sum_{k=1}^m \hat{S}^z_{aj} \hat{S}^z_{bk},
\end{equation}
\noindent is a conserved quantity, $[\hat{H},\hat{S}_T^z]$.

\begin{figure}
\begin{center}
\includegraphics[scale= 0.4]{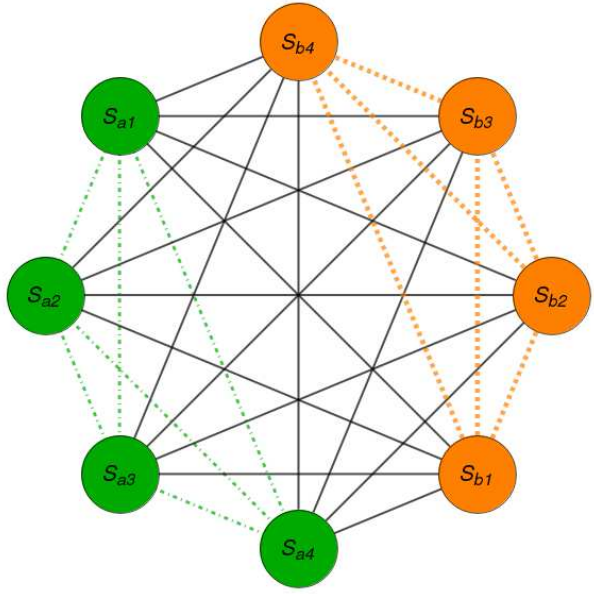}
\caption{The complete graph  $K_8$ showing the interactions between the spins  for $n=m=4$ in the Hamiltonian (\ref{H1spin}). The green dashed-dotted lines stand for interaction between the spins $a$, the orange dashed lines stand for interaction between the spins  $b$, and the black lines stand for interaction between the spins $a$ and the spins $b$.}
\label{gf0}
\end{center}
\end{figure}

We can represent the spins configurations of the Hamiltonian (\ref{H1spin}) as a complete graph $K_s$ (or universal graph) \cite{graph}, where $s = n + m$ is the number of vertices (spins).  In Fig. (\ref{gf0}) we show the $K_8$ graph for the $8$-spins interaction. The number of  edges (interactions) is 

\begin{equation}
\left(\begin{tabular}{c}
s \\
2 \\ 
\end{tabular}
\right) =  \frac{s(s-1)}{2}.
\end{equation}
 This complete graph looks like a spin network in quantum gravity \cite{QG1,QG2} but this full coupling network will in practice always be limited in size, as physical interactions tend to decrease with the distance. 
 \noindent The complete graphs are also the complete $s$-partite graph $K_{s\times 1}$. If $J^z_{aiaj}=J^z_{bkbl}=D_{aj}=D_{bk}=0$ it becames the complete bipartite graph $K_{n,m}$. 
 
\begin{figure}
\begin{center}
\begin{tabular}{cc}
        (a)          &       (b)      \\
\includegraphics[scale=0.3]{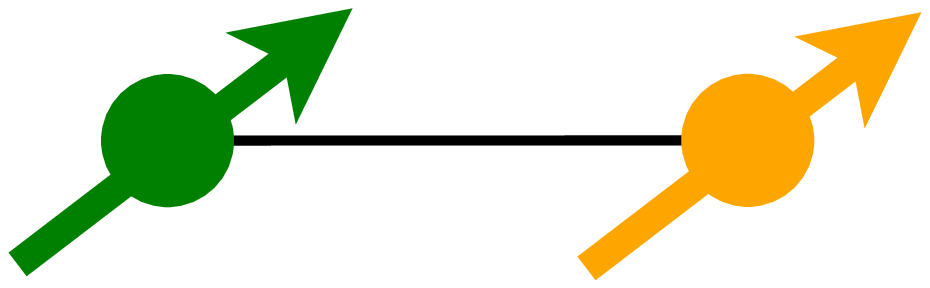} & \includegraphics[scale=0.3]{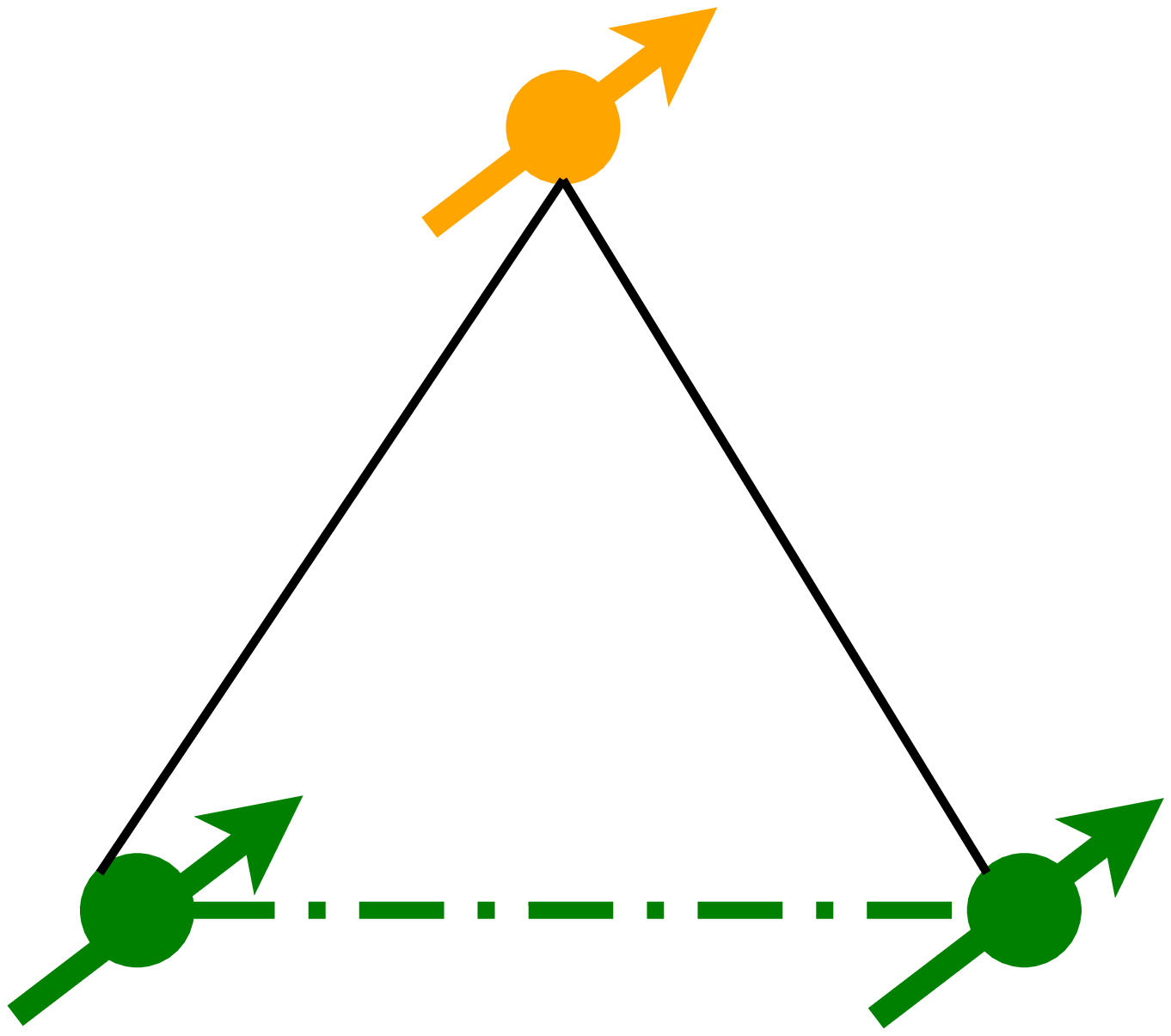} \\
        (c)          &       (d)      \\
\includegraphics[scale=0.3]{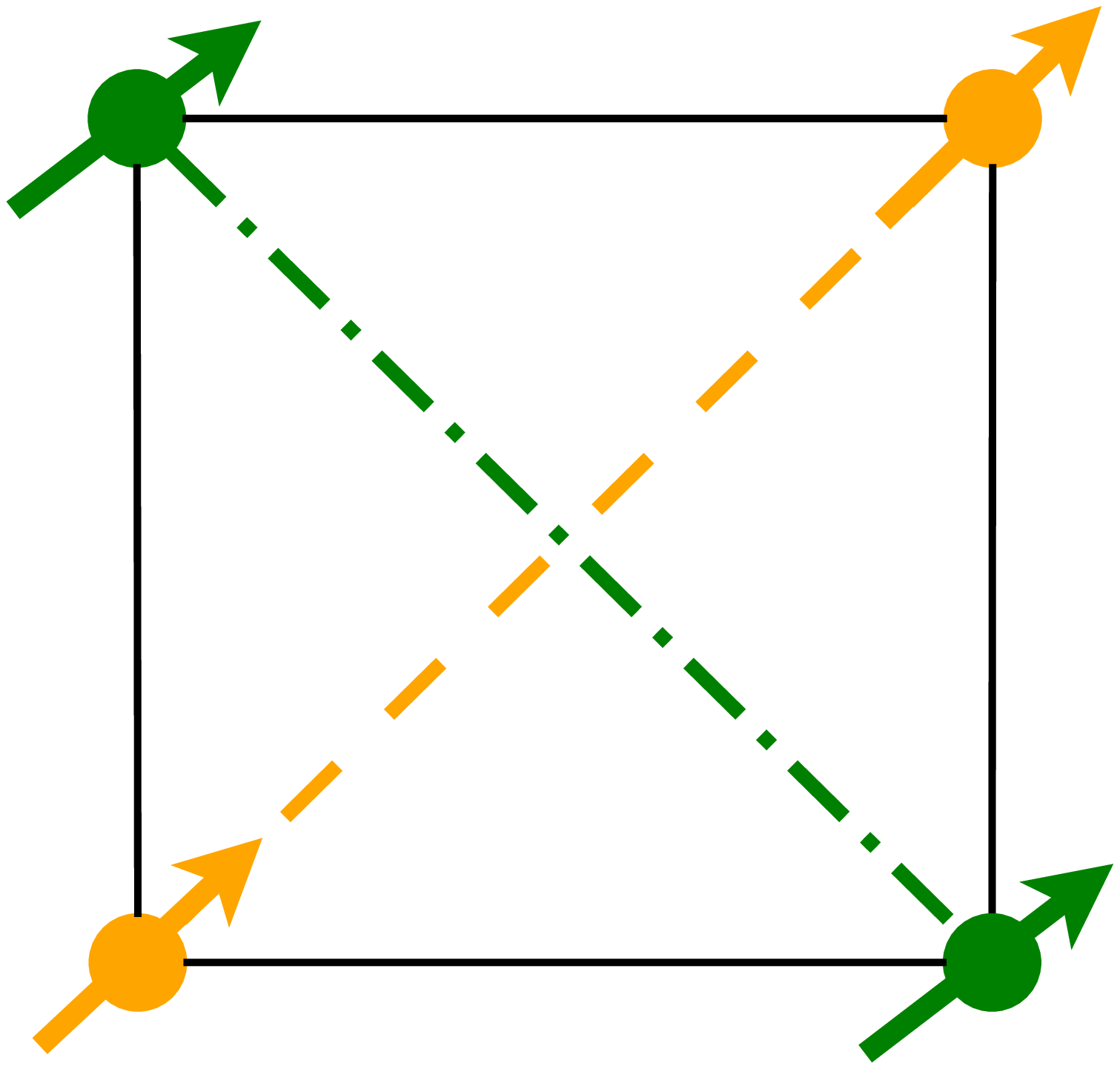} & \includegraphics[scale=0.3]{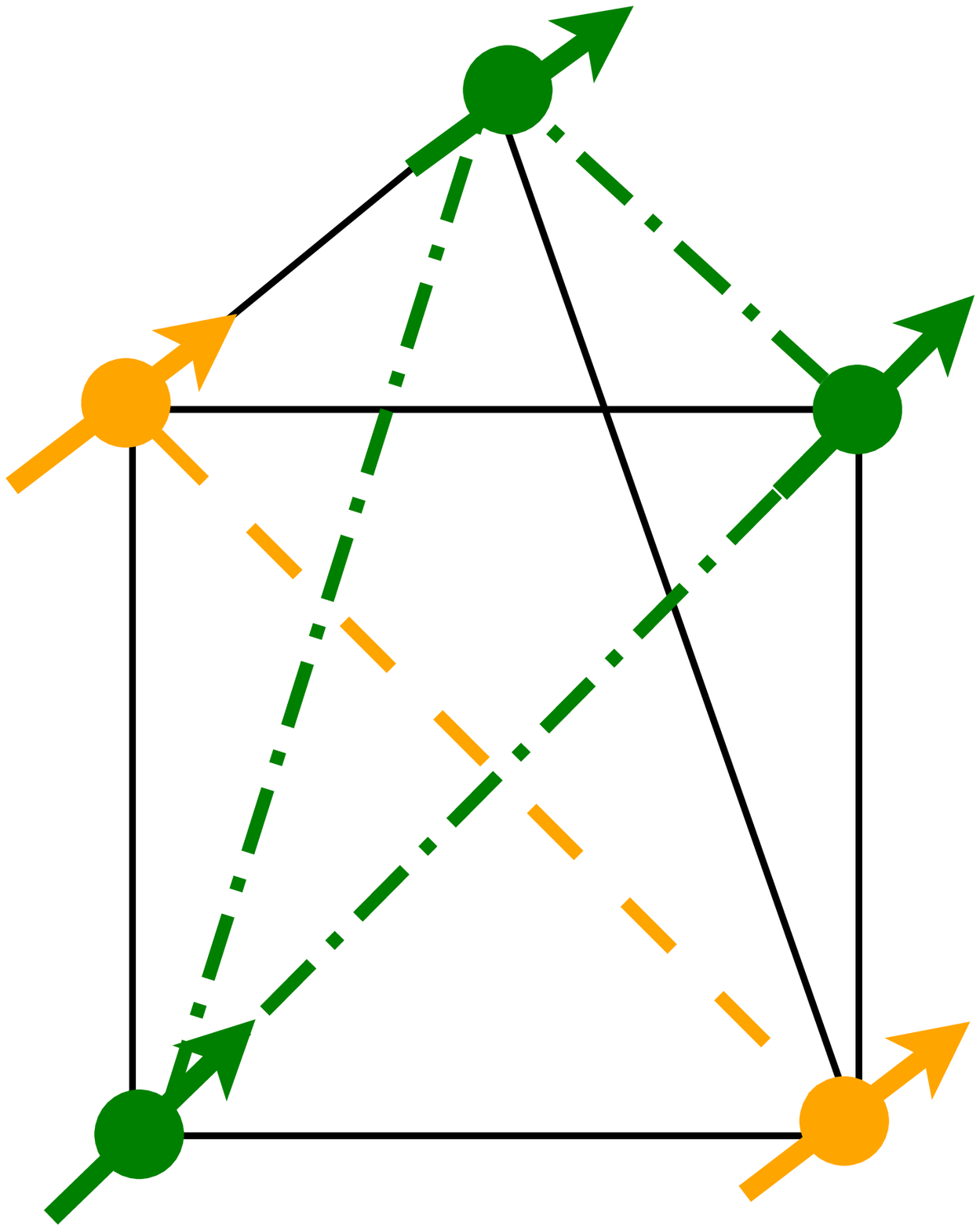} 
\end{tabular}
\caption{Some spin configuration for the generalized two-spins model with $n$ spin $a$ and $m$ spin $b$. The full black lines means interaction between the spins $a$ (green) and the spins $b$ (orange). The green dashed-dotted lines means interaction between the spins $a$ and the orange dashed line means interaction between the spins $b$. $(a)$ 1 spin $a$  and 1 spin $b$.  $(b)$ 2 spin $a$  and 1 spin $b$. $(c)$ 2 spin $a$  and 2 spin $b$. $(d)$ 3 spin $a$  and 2 spin $b$.}
\label{gf1}
\end{center}
\end{figure}

 In Fig.  (\ref{gf1}) we show some of the spin configurations. These  small spin configurations   can be considered as interaction of the nuclear spin of the atoms in a molecule, as in nuclear magnetic resonance, can be considered as plaquettes of a spin ladders or can be considered as  a cell of a 3D lattice.

\newpage

\section{Exact solution}
Now we use the co-multiplication property of the Lax operators to write
\begin{equation}
  \hat{L}(u) = \hat{L}_{1}^{S_a^n}(u + \omega_{a})\hat{L}_{2}^{S_b^m}(u - \omega_{b}).
\label{LHm2}
\end{equation}
\noindent Following the monodromy matrix (\ref{monod}) we can write the operators,

\begin{eqnarray}
\pi(\hat{A}(u)) &=& \left(\gamma_a u\hat{I} +  \gamma_a\omega_{a}\hat{I} - \eta\sum_{j=1}^n \gamma_a\hat{S}^z_{aj}\right)\left(\gamma_b u\hat{I}  -  \gamma_b\omega_{b}\hat{I} - \eta\sum_{k=1}^m \gamma_b\hat{S}^z_{bk}\right) \nonumber \\
          &+& \eta^2\sum_{j=1}^n\sum_{k=1}^m \alpha_{aj}\beta_{bk} \hat{S}^{+}_{aj}\hat{S}_{bk}^{-}, \label{piAn}\\
\pi(\hat{B}(u)) &=& \left(\gamma_a u\hat{I} +  \gamma_a\omega_{a}\hat{I} - \eta\sum_{j=1}^n \gamma_a\hat{S}^z_{aj}\right)\left(-\eta\sum_{k=1}^m \alpha_{bk}\hat{S}^{+}_{bk}\right) \nonumber \\
&+& 
\left(-\eta\sum_{j=1}^n \alpha_{aj}\hat{S}^{+}_{aj}\right)
\left(\rho_b u\hat{I} -  \rho_b\omega_{b}\hat{I} + \eta\sum_{k=1}^m \rho_b\hat{S}^z_{bk}\right),
\label{piBn}
\end{eqnarray}

\begin{eqnarray}
\pi(\hat{C}(u)) &=& \left(-\eta\sum_{j=1}^n \beta_{aj}\hat{S}^{-}_{aj}\right)
\left(\gamma_b u\hat{I} -  \gamma_b\omega_{b}\hat{I} - \eta\sum_{k=1}^m \gamma_b\hat{S}^z_{bk}\right) \nonumber \\
          &+& 
\left(\rho_a u\hat{I} + \rho_a\omega_{a}\hat{I}  + \eta\sum_{j=1}^n \rho_a\hat{S}^z_{aj}\right)\left(-\eta\sum_{k=1}^m \beta_{bk}\hat{S}^{-}_{bk}\right),
\label{piCn}\\
\pi(\hat{D}(u)) &=&  \left(\rho_a u\hat{I} +  \rho_a\omega_{a}\hat{I} + \eta\sum_{j=1}^n \rho_a\hat{S}^z_{aj}\right)\left(\rho_b u\hat{I}  - \rho_b\omega_{b}\hat{I} + \eta\sum_{k=1}^m \rho_b\hat{S}^z_{bk}\right) \nonumber \\ 
&+& \eta^2\sum_{j=1}^n\sum_{k=1}^m \alpha_{bk}\beta_{aj} \hat{S}^{-}_{aj}\hat{S}_{bk}^{+}.
\label{piDn}
\end{eqnarray}

Taking the trace of the operator (\ref{LHm2}) we get the transfer matrix

\begin{eqnarray}
\hat{t}(u) &=&  u^2 \sigma_{ab} \hat{I} + u\sigma_{ab} \Omega_{a-b} \hat{I} - u\eta \Delta_{ab} \hat{S}^z_{T} - \sigma_{ab} \Xi_{a\times b} \hat{I} \nonumber \\   
&+&     \eta \Delta_{ab} \omega_{b}\sum_{j=1}^n \hat{S}_{aj}^z  - \eta \Delta_{ab} \omega_{a}\sum_{k=1}^m  \hat{S}_{bk}^z  + \eta^2 \sigma_{ab} \sum_{j=1}^n\sum_{k=1}^m  \hat{S}_{aj}^z \hat{S}_{bk}^z \nonumber \\
&+& \eta^2 \sum_{j=1}^n\sum_{k=1}^m \left(\alpha_{aj}\beta_{bk} \hat{S}^{+}_{aj}\hat{S}_{bk}^{-} + \alpha_{bk}\beta_{aj} \hat{S}^{-}_{aj}\hat{S}_{bk}^{+} \right),
\label{tgsmb}
\end{eqnarray}
\noindent where $\sigma_{ab} \equiv \gamma_a\gamma_b + \rho_a\rho_b$,  $\Delta_{ab} \equiv  \gamma_a\gamma_b - \rho_a\rho_b$, $\Omega_{a-b} = \omega_{a} - \omega_{b}$, $ \Xi_{a\times b} = \omega_{b}\omega_{b}$.

 From (\ref{C14b}) we identify the conserved quantities of the transfer matrix (\ref{tgsmb}),

\begin{eqnarray}
\hat{\mathcal{C}}_0 
	 & = &  -  \sigma_{ab} \Xi_{a\times b} \hat{I}  +
  \eta \Delta_{ab}\omega_{b}\sum_{j=1}^n\hat{S}_{aj}^z - \eta \Delta_{ab} \omega_{a}\sum_{k=1}^m  \hat{S}_{bk}^z \nonumber \\ &+& \eta^2 \sigma_{ab}\sum_{j=1}^n\sum_{k=1}^m  \hat{S}_{aj}^z \hat{S}_{bk}^z
+ \eta^2 \sum_{j=1}^n\sum_{k=1}^m \left(\alpha_{aj}\beta_{bk} \hat{S}^{+}_{aj}\hat{S}_{bk}^{-} + \alpha_{bk}\beta_{aj} \hat{S}^{-}_{aj}\hat{S}_{bk}^{+} \right), \\
\hat{\mathcal{C}}_1 &=&      \sigma_{ab} \Omega_{a-b} \hat{I} - \eta \Delta_{ab} \hat{S}^z_{T}, \\
\hat{\mathcal{C}}_2 &=& \sigma_{ab} \hat{I}.
\end{eqnarray}

We can rewrite the Hamiltonian  (\ref{H1spin}) using these conserved quantities,
\begin{equation}
 \hat{H} = \xi_0\hat{\mathcal{C}}_0 + \hat{\mathcal{C}}_1(1 + \xi_1\hat{\mathcal{C}}_1)   - \xi_2 \hat{\mathcal{C}}_2,
\label{H2spin} 
\end{equation}
\noindent with the following identification for the parameters
\begin{equation}
\xi_2 = \Omega_{a-b}(1 + \xi_1\sigma_{ab}\Omega_{a-b}) - \xi_0 \Xi_{a\times b},
\end{equation}

\begin{equation}
B^z_{aj} = \eta\Delta_{ab}\left(\xi_0\omega_{b} - 2\xi_1\sigma_{ab}\Omega_{a-b} - 1 \right),
\end{equation}

\begin{equation}
B^z_{bk} = -\eta\Delta_{ab}\left(\xi_0\omega_{a} + 2\xi_1\sigma_{ab}\Omega_{a-b} + 1 \right),
\end{equation}

\begin{equation}
D_{aj}  = \xi_1\eta^2\Delta_{ab}^2,
\end{equation}

\begin{equation}
 D_{bk} = \xi_1\eta^2\Delta_{ab}^2,
\end{equation}

\begin{equation}
J^z_{aiaj} =  2\xi_1\eta^2\Delta_{ab}^2,
\label{diff1}
\end{equation}

\begin{equation}
J^z_{bkbl} = 2\xi_1\eta^2\Delta_{ab}^2,
\label{diff2}
\end{equation}

\begin{equation}
J^z_{ajbk} = \eta^2(\xi_0\sigma_{ab} + 2\xi_1\Delta_{ab}^2),
\end{equation}

\begin{equation}
J^{xy}_{ajbk} = 2\xi_0\eta^2\alpha_{aj}\beta_{bk} = 2\xi_0\eta^2\alpha_{bk}\beta_{aj}.
\label{Jzab}
\end{equation}
\noindent In the Eqs. (\ref{diff1}) and (\ref{diff2}) we consider $i\neq j$ and $k\neq l$.  Using the Eq. (\ref{Jzab}) and the condition $\alpha_{aj}\beta_{aj} = \alpha_{bk}\beta_{bk}$ $(\gamma_a\rho_a = \gamma_b\rho_b)$ we get the following relation for the parameters $\sigma_{ab}$ and $\Delta_{ab}$,

\begin{equation}
\sigma_{ab} = \frac{\rho_b}{\rho_a}(\gamma_b^2 + \rho_a^2),    \qquad   \Delta_{ab} =   \frac{\rho_b}{\rho_a}(\gamma_b^2 - \rho_a^2).
\end{equation}
\noindent The $\xi_0$, $\gamma'$s and $\rho'$s parameters are arbitrary, but different of zero. The $\xi_1$ parameter can be zero and in this case we take out the terms with anisotropy parameter, the term with the interactions between the spins $a$ and the term with interactions between the spins  $b$. We also can use $\gamma_b^2 = \rho_a^2$ to cancel  these interactions and turn off the $B$-field. The $\omega'$s parameters are completely arbitrary.

The Hamiltonians (\ref{H1spin}) and (\ref{H2spin}) are related to the transfer matrix (\ref{tgsmb}) by the equation,

\begin{equation}
 \hat{H} = \xi_0\hat{t}(0) + \hat{t}'(0)[1 + \xi_1 \hat{t}'(0)] - \xi_2 \hat{t}''(0),
\label{H4spin} 
\end{equation}
\noindent where the prime symbol $( ' )$ stand for  derivatives of $\hat{t}(u)$.

We use as pseudo-vacuum the product state, 

\begin{equation}
|0\rangle_{ajbk} = \left(\bigotimes_{j=1}^n|\uparrow_z\rangle_{aj}\right)\otimes \left(\bigotimes_{k=1}^m|\uparrow_z\rangle_{bk}\right),
\end{equation}
\noindent with $|\uparrow_z\rangle_{aj}$ denoting the vacuum state for the spins $aj$ and $|\uparrow_z\rangle_{bk}$ denoting the vacuum state for the spins $bk$, for $j=1,\ldots, n$ and $k=1\ldots, m$. For this pseudo-vacuum we can apply the algebraic Bethe ansatz method in order to find the Bethe ansatz equations (BAEs),

\begin{eqnarray}
\frac{\gamma_a\gamma_b\left( v_i +  \omega_{a} - \eta M^z_{a}\right) \left( v_i  -  \omega_{b} - \eta M^z_{b}\right)}{\rho_a\rho_b\left( v_i + \omega_{a} + \eta M^z_{a}\right)\left( v_i  -  \omega_{b} + \eta M^z_{b}\right)} & = & \prod_{j \ne i}^{N}\frac{v_{i}-v_{j}-\eta}{v_{i}-v_{j}+\eta}, \;\;\;\;\;  i,j = 1,\ldots , N, \nonumber\\
\label{BAE2}
\end{eqnarray}
\noindent where

\begin{equation}
M^z_{a} = \sum_{j=1}^n m^z_{aj},  \qquad  M^z_{b} = \sum_{k=1}^m m^z_{bk},
\end{equation}
\noindent are the total magnetic moment in the $z$-direction of the respective spins $a$ and $b$. 

If we choose $\omega_a = \omega_b = 0$,  $\gamma_b = \rho_a$ and $M^z_{a} = M^z_{b} \equiv M^z$ we get the BAE's

\begin{eqnarray}
\left(\frac{ v_i  - \eta M^z}{ v_i +\eta M^z}\right)^2 & = & \prod_{j \ne i}^{N}\frac{v_{i}-v_{j} - \eta}{v_{i}-v_{j} + \eta}, \;\;\;\;\;  i,j = 1,\ldots , N.
\label{BAE3}
\end{eqnarray}

The eigenvectors \cite{Bethe-states}  $\{ |v_1,v_2,\ldots,v_N\rangle \}$ of the Hamiltonian (\ref{H1spin}) or (\ref{H2spin}) and of the transfer matrix (\ref{tgsmb}) are 

\begin{equation}
|\vec{v}\rangle \equiv  |v_1,v_2,\ldots,v_N\rangle = \prod_{i=1}^N \pi(\hat{C}(v_i))|0 \rangle,
\end{equation}
\noindent and the eigenvalues of the Hamiltonian (\ref{H1spin}) or (\ref{H2spin}) are,

\begin{eqnarray}
E(\{ v_i \}) & = &   \xi_0\; \gamma_a\gamma_b \left( v_i+ \omega_{a} - \eta  M^z_{a}\right) \left( v_i  -  \omega_{b} - \eta  M^z_{b}\right) \prod_{i=1}^{N}\frac{v_{i} + \eta}{v_{i}} 
  \nonumber \\
  &+&  \xi_0\; \rho_a\rho_b \left( v_i +  \omega_{a} + \eta M^z_{a}\right)\left( v_i  - \omega_{b} + \eta M^z_{b}\right)\prod_{i=1}^{N}\frac{v_{i}  - \eta}{v_{i}} \nonumber \\
   &-& \eta\Delta_{ab}\left(1 +  2\xi_1\;\sigma_{ab}\Omega_{a-b}\right)\left(M^z_{a} + M^z_{b}\right) \nonumber \\ 
   &+& \xi_1\;\eta^2 \Delta_{ab}^2 \left[(M^z_{a})^2 + 2 M^z_{a} M^z_{b}+ (M^z_{b})^2 \right]  
   + \xi_0\sigma_{ab}\Xi_{a\times b},
\end{eqnarray}
\noindent where the $\{v_i\}$ are solutions of the BAEs (\ref{BAE2}).

\section{Summary}
We have solved a generalized two-spins model by the algebraic Bethe ansatz method using the $gl(2)$-invariant $R$-matrix and a multi-spins Lax operator. In this generalized two-spins model we are  considering two spins: $a$ and $b$. Each spin interacts with all the others spins. The interactions are by the Heisenberg spins exchange. We are also considering a $B$-field in the $z$-direction and a term for Haldane single spin anisotropy. We can represent all spins interaction as a complete graph $K_s$, where $s$ is the total number of spins $a$ plus the total number of spins $b$ and we use this graph to calculate the number of interactions. We can take out the interaction between the  spins $a$, the interaction between the spins $b$  and the single spin anisotropy simultaneously to get a complete bipartite graph $K_{n,m}$. We also can turn off the externals $B$-fields.

\section*{Acknowledgments}
The author acknowledge the Physics Department of the Universidade Federal de Sergipe for the  support.

\end{document}